
\input phyzzx

\def\SCIPP{\centerline {\it Santa Cruz Institute for Particle Physics}
  \centerline{\it University of California, Santa Cruz, CA 95064}}
\overfullrule 0pt
\titlepage
\Pubnum{SCIPP 92/27}
\date{June, 1992}
\pubtype{ T}     
\title{{
Problems of Naturalness:  Some Lessons from String Theory
}
\foot{Invited Talk Presented at the Cincinnati Symposium in Honor
of the Retirement of Louis Witten}
\foot{Work supported in part by the U.S. Department of Energy.}}
\author{Michael Dine
}
\address{}
\SCIPP
\vskip.5cm
\vbox{
\centerline{\bf Abstract}
We consider some questions of naturalness which arise when
one considers conventional field theories in the presence of
gravitation:  the problem of global symmetries, the strong CP
problem, and the cosmological constant problem.  Using string
theory as a model, we argue that it is reasonable to postulate
weakly broken global discrete symmetries.
We review the arguments that gravity is likely to spoil the Peccei-Quinn
solution of the strong CP problem, and update earlier analyses
showing that discrete symmetries can lead to axions with suitable
properties.  Even if there are not suitable
axions, we note that
string theory is a theory in which
CP is spontaneously broken and $\theta$ in principle calculable.
$\theta$ thus might turn out to be small along lines suggested
some time ago by Nelson and by Barr.}

\parskip 0pt
\parindent 25pt
\overfullrule=0pt
\baselineskip=18pt
\tolerance 3500

\endpage
\pagenumber=1

\bigskip

\chapter{Introduction}

 Rather than deal here, as other speakers will, directly
with the difficult questions raised by quantum gravity,
I would like to focus on some questions of naturalness which
Einstein's theory raises.  The three which will concern
us here are:  the cosmological constant; the problem of
symmetries (both continuous and discrete) and, related to the
second, the strong CP problem.  In considering these questions,
we will use string theory as a guide.  In doing this, I am
not assuming that string theory necessarily describes the
real world, but rather that characteristics of string theory might
plauisbly be shared by any ultimate theory of nature.

The cosmological constant problem arises already if we consider
(semi)-classical gravity coupled to quantum fields.  At one loop,
for example, in field theory one has a contribution to the vacuum
energy
$$E_o = \Lambda =  \sum_i {\pm} {1/2} \int^{\mu} {d^3 k \over (2 \pi^3)}
\sqrt{\vec k^2 + m_i^2}\eqn\cosmocon$$
Here the sum runs over all physical helicity states in the theory;
the $\pm$ refers to bosons and fermions, respectively.
Generically, the result is
$$E_o \sim \mu^4 \eqn\cosmocon2$$
If $\mu \sim M_P$, this corresponds to a cosmological constant more than
$120$ orders of magnitude larger than the observational limit.  In the
presence of supersymmetry,
the leading divergence cancels between bosons and fermions and
one might hope to find $\mu$ of order
the supersymmetry breaking scale, perhaps as small as $10^2$ GeV.\foot{
Aficianados of hidden sector supergravity models will object that
$\mu \sim 10^{11}~ GeV$ is more reasonable; I am just describing the
best one can hope for.}
\REF\rohm{R. Rohm, Nucl. Phys. {\bf B237} (1984) 553.}
This is a big
improvement, but not nearly good enough.
Of course, in field theory the cosmological
constant is not calculable, and it is not clear we are asking a
physically meaningful question.  However, in string theory the
cosmological constant {\it is} calculable; whenever supersymmetry
is broken one finds it is large.\refmark{\rohm}  So, with regards
to this problem, string theory seems to offer no miracles:  we will
still need to search for some deeper explanation.

Our second topic has to do with symmetries.
The notion of an exact global symmetry is always a troubling one; it is
particularly so in the presence of gravity.  Global quantum numbers can
disappear in black holes; wormholes, if they are relevant, will
generate symmetry-violating operators.  Gauge symmetries, on the
other hand, enjoy a different status, and are expected to survive
quantum gravity effects.  Both these statements apply not only to
\REF\krauss{L. Krauss and F. Wilczek, Phys. Rev. Lett. {\bf
62}, 1221 (1989).}
continuous symmetries, but to discrete symmetries as well.
Krauss and Wilczek have stressed that gauged discrete symmetries
should survive quantum gravity effects.\refmark{\krauss}
The simplest example of such a symmetry is provided by a spontaneously
broken $U(1)$ gauge symmetry.  Suppose, for example, one has two
scalar fields, $\phi$, with $q=2$, and $\chi$, with $q=1$.
An expectation value for $\phi$ leaves over the symmetry $\chi
\rightarrow -\chi$.

\REF\bdm{T. Banks, L. Dixon, D. Friedan and E. Martinec, Nucl.
Phys. {\bf B299} (1988) 613.}
String theory lends weight to these views.  It is not difficult to
prove that, even at tree level, the theory possesses no
unbroken, continuous global symmetries.\refmark{\bdm}
The strategy is to show
that any such symmetry implies the existence of a conserved world
sheet current, which in turn implies the presence of a massless
\REF\wittendiscrete{E. Witten, Nucl. Phys. {\bf B258} (1985) 75.}
vector particle.  Discrete symmetries
do frequently arise in compactifications of
string theory.\refmark{\wittendiscrete}  In many cases,
these can be interpreted  as relics of higher dimensional gauge
and general coordinate invariance, i.e. as gauge symmetries.
It has been widely speculated that all discrete symmetries in string
theory of this kind; we will have more to say about this later.

The third problem we have mentioned is the strong CP problem.
QCD possesses an additional parameter, $\theta$, which enters
the lagrangian through the term
$${\cal L}_{\theta} = \theta {g^2 \over 16 \pi^2} \int d^4 x
F \tilde F \eqn\thetaterm$$
{}From the limits on the neutron electric dipole moment, one knows
\REF\crewther{R.J. Crewther, P. Di Vecchia, G. Veneziano and E.
Witten, Phys. Lett. {\bf 88B} (1979) 123.}
that $\theta < 10^{-9}$.\refmark{\crewther}
Two solutions to this problem have been widely considered.  Perhaps
\REF\pq{R.D. Peccei and H.R. Quinn, Phys. Rev. Lett. {\bf 38}
(1977) 1440; Phys. Rev. {\bf D16} (1977) 1791;
S. Weinberg, Phys. Rev. Lett. {\bf 40} (1978) 223;
F. Wilczek, Phys. Rev. Lett. {\bf 40} (1978) 279.}
the most popular is the ``axion."\refmark{\pq}  Here one postulates that
the classical lagrangian posseses a global $U(1)$ symmetry, the
``Peccei-Quinn" symmetry, under which there is a massless field,
the axion, which transforms non-linearly
$$a(x) \rightarrow a(x) + \delta \eqn\nonlinearlaw$$
The axion is assumed to couple to $F \tilde F$ as
$${\cal L}_a = {N g^2 \over 16 \pi^2} \int d^4 x(\theta
+{ a \over f_a}) F \tilde F\eqn\axioncoupling$$
$f_a$ is the axion decay constant.
QCD effects can then be shown to generate a potential for the axion,
$$V(a) \approx -m_{\pi}^2 f_{\pi}^2 cos({a \over f_A} + \theta)
\eqn\axionpotential$$
The minimum of this potential clearly occurs when $\theta_{eff}
= {a \over f_a} + \theta = 0$.

But the whole idea of the Peccei-Quinn symmetry is quite puzzling.  Not
only is one postulating a global symmetry, but a symmetry which is
necessarily broken explicitly!  String theory offers some insight
into this question.  Indeed, E. Witten pointed out early on
\REF\wittenaxion{E. Witten, Phys. Lett. {\bf B149} (1984) 359.}
that string theory exhibits symmetries of precisely this
type.\refmark{\wittenaxion}  This can be understood in a number of ways.
For example, if one compactifies the heterotic string to four dimensions,
there is a two-index antisymmetric tensor, $B_{\mu \nu}$,
$\mu, \nu=0, \dots 3$.  The corresponding gauge-invariant field
strength is $H_{\mu \nu \rho} = \partial_{\mu}B_{\nu \rho} + CS$,
where $CS$ denotes the Chern-Simons term.  Such an antisymmetric
tensor is equivalent to a scalar field;
$$\partial_{\mu} a \propto \epsilon_{\mu \nu \rho \sigma} H^{\nu \rho
\sigma}\eqn\duality$$
Because in perturbation theory the low energy effective lagrangian
must be written in terms of $H$, no non-derivative couplings of $a$
appear, so in perturbation theory the lagrangian is symmetric
under $a \rightarrow a + \delta$.
\foot{Alternatively, this statement can be understood in terms of string
vertex operators.  Axion emission is described
by $\int d^2 \sigma \epsilon_{\mu \nu}(k) \epsilon^{\alpha \beta}
\partial_{\alpha} x^{\mu} \partial_{\beta} x^{\nu} e^{i k \cdot x}$.
At zero momentum, this becomes the integral of a total divergence.}

Thus one has a symmetry to all orders of perturbation
theory, broken by effects of order $e^{-1 / g^2}$
\REF\shenker{S. Shenker, Rutgers preprint RU-90-47 (1990).}
(perhaps $e^{-1 / g}$\refmark{\shenker}).
This sounds like precisely what one needs to solve the strong
CP problem.  So perhaps it is not so unreasonable, in general,
to postulate such symmetries.

What about the possibility that CP is spontaneously violated,
with vanishing bare $\theta$?  Below we will argue that
in the heterotic string theory, $CP$ is indeed conserved at a
fundamental level; all observable $CP$ violation is necessarily
spontaneous and, in principle, calculable.

We now take up each of the issues raised here in more detail.

\chapter{Discrete Symmetries}

We have argued that gauged discrete symmetries are safe,
i.e. they are unbroken by gravity.  We have also remarked that such symmetries
arise in field theory and are quite common in string theory.
We will now show that, unlike continuous symmetries, {\it approximate
global}
discrete symmetries also arise in string theory.

To motivate our treatment of this subject, consider the problem of
anomalies.
We are used to the notion that continuous gauge symmetries
\REF\thooft{G. 't Hooft, Phys. Rev. Lett. {\bf 37} (1976),
Phys. Rev. {\bf D14} (1976) 3432.}
\REF\ir{L. Ibanez and G. Ross, Phys. Lett. {\bf 260B} (1991) 291;
Nucl. Phys. {\bf B368} (1992) 3.}
\REF\trivedi{J. Preskill, Sandip Trivedi, F. Wilczek and M. Wise,
Nucl. Phys. {\bf B363} (1991) 207.}
\REF\wittenprivate{E. Witten, private communication c. 1983.}
should be free of anomalies.  What about discrete symmetries?
That anomalies can arise in discrete symmetries can be understood
by considering instantons in an effective low energy theory.
Instantons generally give rise to effective operators which
break symmetries; 't Hooft showed long ago,
for example, that instantons of the
electroweak theory generate an effective interaction
which breaks both baryon and lepton numbers.
The effective interactions generated by instantons can also
violate discrete symmetries.  For a gauge symmetry, such
a breaking signals an {\it inconsistency}, and can be viewed as an
anomaly.\refmark{\ir,\trivedi,\wittenprivate}  One can attempt to
understand discrete anomalies by embedding discrete symmetries
in continuous ones.\refmark{\ir}
However, this leads to constraints which
depend on the quantum numbers of massive fields.  The only
constraints on discrete symmetries which involve exclusively
properties of light fields can be understood in terms of
instantons
\REF\banksdine{T. Banks and M. Dine, Phys. Rev. {\bf D45} (1992)
424.}
in the effective low energy theory.\refmark{\banksdine,\ir}

What about string theory?  If we assume that all discrete symmetries
in string theory are gauge symmetries, it is natural to
ask whether discrete anomalies ever arise for modular-invariant
compactifications.  If one found compactifications with
such anomalies, they would
be inconsistent.  Such a situation
would be remeniscient of global anomalies.  It could be
quite dramatic, representing a new, non-perturbative
consistency condition on string
compactifications.  A priori, I don't know of an argument
that this can not occur;  indeed,
I know of no general argument that other
types of global anomalies (e.g. SU(2) anomlies)
do not occur.

In fact, study of various compactifications
\REF\discrete{Several examples of the phenomena which we now
describe were presented in ref. \banksdine.  $1000$'s
more have been studied by D. MacIntire (SCIPP preprint in preparation).}
quickly yields numerous examples of anomalies!\refmark{\discrete}
However, in all the cases which have been studied  to date,
one can cancel these anomalies in the following way.
\foot{The possibility that anomalies in discrete symmetries
might be cancelled by a Green-Schwarz mechanism was noted in
ref. \ir.}
The axion, $a$, couples to all of the low energy gauge groups:
$$ {g^2 \over 16 \pi^2 f_a} \sum a(x) F^{(i)} \tilde F^{(i)}
\eqn\allgroups$$
It turns out that one can always cancel all of the anomalies by
assigning to the axion a non-linear transformation law of the
form
$${a \over f_a} \rightarrow {a\over f_a} + \beta \eqn\discretetrans$$
for some number $\beta$.
To understand what is going on here, note that the instanton effective
action is typically something of the form
$$\psi \psi \dots \psi e^{i a/f_a} e^{-8 \pi^2/g^2}\eqn\instantoneffect$$
So the phase rotation of the fermions is compensated by the
shift in the axion field.  This result is highly non-trivial (typically
several anomalies are being taken care of by one such shift).
It almost surely indicates that the symmetries are not, in fact,
anomalous.  So far only rather special classes of models have been
examined, so that while I suspect that this is a general result,
it is by no means certain.  In any case, there is still no evidence for
the existence of any new, (independent) consistency condition beyond
those which hold in perturbation theory.


However, from these studies we learn something suprising:  string
theory possesses {\it global} discrete symmetries which are valid
to any order of perturbation theory and broken only non-perturbatively.
For while the non-anomalous symmetry in all of these cases
is {\it spontaneously} broken by the non-linear transformation law
of the axion, the original, anomalous symmetry is good to all orders,
being broken only non-perturbatively.  If we adopt the view that
phenomena which occur in string theory can plausibly occur in any
ultimate theory, this means that it is reasonable to postulate
approximate global symmetries in a low energy theory.  Such
symmetries have been suggested for many reasons, such as avoiding
flavor changing neutral currents in multi-Higgs theories and proton
decay in superysmmetric theories, for understanding the fermion
mass matrix, and (see below) for understanding the strong CP problem.

\chapter{Strong $CP$}

\section{Is CP Spontaneously Broken in String Theory?}

\REF\strominger{A. Strominger and E. Witten, Comm. Math. Phys. {\bf
101} (341) 1985.}
In perturbation theory,
$CP$ is {\it conserved} in string theory.\refmark{\strominger}
\REF\douglas{M. Douglas and S. Shenker, Nucl. Phys. {\bf B355}
(1990) 635.}
One might ask whether this is true non-perturbatively.
After all, in field theory,
$\theta$ is a non-perturbative
parameter which violates CP.
It has been suggested that string theory might possess
similar non-perturbative parameters.\refmark{\douglas}
If some of these are $CP$-violating,
they might give rise to $\theta$ parameters in the low energy theory.
\REF\cpgauge{K. Choi, D. Kaplan and A. Nelson, UCSD preprint PTH 92-11
(1992);
M. Dine, R. Leigh and D. MacIntire, SCIPP preprint SCIPP 92/16
 (1992).}
However, it turns out that one can argue that $CP$ is a gauge symmetry
in string theory.\refmark{\cpgauge}
\foot{When I presented this talk in Cincinnati, I was not sure of this
statement, and only mentioned it as a possibility.  I offered in
addition some alternative arguments for absence of $\theta$ parameters.}
This means that there can be no such CP-violating parameters,
since these would correspond to an {\it explicit} breaking
of the symmetry.

As a result,
if string theory describes nature, CP must be spontaneously broken
and $\theta_{QCD}$ is calculable.  This breaking might arise at
$M_p$ (e.g. through expectation values for CP-odd moduli)
or at lower scales (e.g. through vev's for some matter fields).
In either case, one expects that
generically $\theta$ will be large, proportional to
other $CP$-violating phases needed to explain the features of the
$K$-meson system.  However, in field theory, it is known that one
\REF\nelson{A. Nelson, Phys. Lett. {\bf 136B} (1984) 387;
S.M. Barr, Phys. Rev. Lett. {\bf 53} (1984) 329; Phys.
Rev. {\bf D30} (1984) 1805; P.H. Frampton and T.W. Kephart,
Phys. Rev. Lett. {\bf 65} (1990) 1549.}
can sometimes arrange things so that $\theta$ is small.\refmark{\nelson}
\REF\dine{M. Dine, SCIPP preprint to appear.}
Preliminary investigation (to be described in ref. \dine) indicates
that certain ``string inspired models" can accomplish this.
In particular, in a class of models, discrete symmetries insure
that $CP$ is spontaneously
violated at an ``intermediate scale", $M_{INT}$, of order $10^{11}$ GeV,
with $\theta$ of order $M_{INT} \over M_p$ (times coupling
constants).  Moreover, in these models,
the low energy theory is supersymmetric, but the only CP violation
lies in the KM phase and $\theta$.


\section{Accidental Axions in String Theory}

Alternatively, one can explore axion solutions to the strong CP problem
in string theory.  There are, however, two potential problems
with the stringy axion, $a$, which we have described above.
First, in many compactifications of string theory, there is more than
one strongly interacting gauge group; it is necessary to
have at least one axion for each group.  Second, even if $QCD$ is
the only strong group, the decay constant, $f_a$, is
\REF\axioncosmology{J. Preskill, M. Wise and F. Wilczek,
Phys. Lett. {\bf 120B}(1983) 127; L. Abbott and P. Sikivie,
Phys. Lett. {\bf 120B} (1983) 133; M. Dine and W. Fischler,
Phys. Lett. {\bf 120B} (1983) 137.}
a number of order $M_P$.  This contradicts cosmological
bounds, which give $f_a < 10^{12}$ GeV.\refmark{\axioncosmology}
However, one might choose to ignore these limits; there are
a number of possible loopholes.  For example, these analyses
assume that there is no entropy generation after the QCD phase transition.
However, plausible models exist in which there is such entropy generation,
\REF\raby{J. Cline and S. Raby, Phys. Rev. {\bf D43} (1991) 1381.}
and yet an adequate baryon density is generated.\refmark{\raby}
These arguments also assume that the initial value of the axion field
in the observable universe is simply a random number; in that case,
for such a large $f_a$, only one universe in $10^3$ has a sufficiently
small initial $\theta$.
\REF\linde{A.D. Linde, Phys. Lett. {\bf 259B} (1991) 38.}
But Linde has pointed out that the size of the initial $\theta$ may
be correlated with primordial density fluctuations.  Only those
regions with small enough $\theta$, in this view, might resemble
ours.\refmark{\linde}  Thus a rather mild application of the
\REF\weinberg{S. Weinberg, Phys. Rev. Lett. {\bf 59} (1987)
2607; Rev. Mod. Phys. {\bf 61} (1989) 1.}
anthropic principle (the ``weak anthropic principle"\refmark{\weinberg})
might solve the problem.
You may not wish to take any of these possibilities too seriously;
however, one should be aware that the cosmological axion limit
rests on certain assumptions which may not be true.


For now, though, let us take the cosmological limit seriously, and
ask how $f_a \sim 10^{11}$ might arise.  We could, of course, simply
postulate that there is another fundamental scale, and the axion
arises in a manner similar to the string axion.  Such an assumption
is certainly troubling, however, and there is no reason to think such
a scale should arise in string theory.  Alternatively, the Peccei-Quinn
symmetry might arise by accident, in the same way that baryon and lepton
number arise in the standard model.  Such an accident, however,
would be quite startling if one simply assumes that gravity
\REF\wise{H. Georgi, S. Glashow and M. Wise, Phys. Rev. Lett.
{\bf 47} (1981) 402.}
\REF\march{M. Kamionkowski and J. March-Russell, Phys. Lett.
{\bf 282B} (1992) 137;; R. Holman {\it et al.},
Phys. Lett. {\bf 282B} (1992) 132;
S.M. Barr and D. Seckel, Bartol preprint BA-92-11.}
generates all operators consistent with the various local symmetries
of the theory.  The problem is that in order that the axion tune
$\theta$ to the required precision, it is necessary that the leading
operators which violate the symmetry be of very high dimension.
This point was already raised in passing by Georgi, Glashow
and Wise\refmark{\wise}
More recently, it has been discussed
in a general and quantitative
fashion by several authors.\refmark{\march}
To gain some appreciation of the difficulty,
suppose that the lowest dimension,
 gauge-invariant operator which violates the symmetry
is ${\cal O}^{(4+n)}$, of dimension
$4+n$.  Then the leading symmetry-violating term which can occur in
a low-energy effective field theory is
$${\cal L}_{SB} = {\gamma \over M_P^n} {\cal O}^{(4+n)}\eqn\lsb$$
where $\gamma$ is a dimensionless coupling constant.
On dimensional grounds, this gives rise to a linear term in the
axion potential,
$$V_{SB} \propto \gamma{f_a^{n+3} \over M_P^n} a(x)$$
Since
$$m_a^2 \sim {m_{\pi}^2 f_{\pi}^2 \over f_a^2}\eqn\axionmass$$
the resulting shift in $\theta$ is
$$\delta \theta = {\delta a \over f_a} \sim {\gamma \over m_{\pi^2}
f_{\pi}^2} {f_a^{n+4} \over M_P^n}~<~ 10^{-9}\eqn\thetashift$$
For $f_a = 10^{11}$, this gives $n>6$ (i.e. the symmetry-violating
operator must at least be of dimension $12$!)  If $f_a = 10^{10}$,
things are slightly better; one needs to suppress all operators
of dimension less than $9$.

The lesson of all this is that if one wants a Peccei-Quinn
symmetry to arise by accident, one must forbid operators
up to very high dimensions.  How might such a thing occur?
The authors of refs. \march\ noted that
with a sufficiently complicated continuous gauge symmetry,
one could indeed suppress operators of very high dimension.
However, by their own admission, the resulting models were not
particularly beautiful.

\REF\shafi{G. Lazarides, C. Panagiotakopoulos and Q. Shafi,
Phys. Rev. Lett. {\bf 56} (1986) 432.}
\REF\ross{J. Casas and G. Ross, Phys. Lett. {\bf 192B} (1987) 119.}
In light of our earlier discussion,
it is natural to ask
how easily discrete symmetries can accomplish the same
objective.  In fact, in the framework of string theory,
this question was asked some time ago by Lazarides et
al\refmark{\shafi}
and by Ross and Casas.\refmark{\ross}  The latter authors also attempted
to estimate how large a $\theta$ would be induced by
higher-dimension operators which violated the Peccei-Quinn
symmetry, in precisely the spirit described above (we will see,
however, that they failed to consider the most dangerous class
of operators).
Before reviewing these models, however, it
is perhaps useful to illustrate just how
powerful discrete symmetries
are in this respect by considering theories in
which the Peccei-Quinn symmetry is dynamically broken by fermion
condensates.\foot{This has been noted
independently, and much earlier, by A. Nelson (unpublished).}
As an example, consider a theory with (unbroken)
gauge group (in addition to the standard
model gauge group) $SU(4)_{AC}$ ($AC$ is for ``axi-color"), with
scale $\Lambda_{AC} \sim f_a$.  In addition to the usual quarks
and leptons, we suppose that the theory contains additional fields
$Q$ and $\bar Q$, transforming as $(4,3)$ and $(\bar 4, \bar 3)$ under
$SU(4)_{AC} \times SU(3)_c$, and fields ${\cal Q}$ and
$\bar {\cal Q}$ transforming as a $(4, 1)$ and a $(\bar 4, 1)$.
Now suppose that the model possesses a discrete symmetry (gauged
or global) under which
$$Q \rightarrow \alpha Q~~~~~~~~{\cal Q} \rightarrow \alpha {\cal Q}$$
where $\alpha = e^{2 \pi i \over N}$; all other fields are neutral.
If, for example,
$N=3$, the lowest dimension chirality-violating operators
one can write are of the form $(\bar Q Q)^3$, which is dimension $9$;
suppression of still higher dimension operators is achieved by choosing
larger $N$.  In this theory, the would-be PQ symmetry is
$$Q \rightarrow e^{i \omega Q}~~~~~~~~{\cal Q} \rightarrow\
e^{-3 i \omega} {\cal Q} \eqn\pqtrans$$
This symmetry has no $SU(4)$ anomaly, but it does have a QCD anomaly.
One expects that this symmetry will be broken by the
condensates
$$<\bar Q Q> \sim <\bar {\cal Q} {\cal Q}> \sim f_a^3\eqn\condensates$$
This gives rise to an axion with decay constant $f_a$, which solves
the strong CP problem.


Let us turn now to the ideas of Lazarides et al and of Casas and
Ross.  In particular, we will develop a variant of the model
of the latter authors.
Of course, it is not presently clear how string theory might
describe the real world, so we will view this model as
``string inspired," in that it shares features common
to a class of compactifications.  We will have to assume,
also, some structure of soft supersymmetry breaking.
Having said that, it should be stressed that models of this
kind have a major virtue:  the axion decay constant is naturally
of order $M_{INT}=\sqrt{M_{W} M_P}$, i.e.
within the allowed axion window.  Our only truly new point,
beyond those made in ref. \ross,
will be that there are operators beyond those
considered by these authors
which one must eliminate if one is to insure sufficiently
small $\theta$.

Consider a theory with unification in the gauge
group $E_6$, with $E_6$ broken to a rank $6$ group at the unification
scale; this is the structure which emerges from conventional
\REF\gsw{M. Green, J. Schwarz and E. Witten, {\it Superstring Theory},
Cambridge University Press, New York, 1986.}
Calabi-Yau compactification.\refmark{\gsw}
Ordinary matter fields
will be assumed to arise from $27$'s of
$E_6$.
Casas and Ross assume that the theory possesses a $Z_3 \times Z_2$
symmetry.  The $27$ contains two standard model singlets, which
we denote by $S$ and $N$.  These authors suppose that there
are two fields with the quantum numbers of $S$, $S_i$,
and two fields with the quantum numbers of $\bar S$, $\bar S_i$.
Under $Z_3 \times Z_2$, these fields transform as follows:
$$S_1 \rightarrow -\alpha S_1~~~~~\bar S_1 \rightarrow - \alpha \bar
S_1~~~~~~~~~~\alpha=e^{2 \pi i \over 3}\eqn\stransform$$
while $S_2$ and $\bar S_2$ are invariant.
The leading terms allowed in the superpotential are
$$W = {a \over M_p} S_2^2 \bar S_2^2 + {b \over M_p^3} S_1^3 \bar S_1^3
+ {c \over M_p^9} {S_1^6 \bar S_2^6} + {d \over M_p^9}{\bar S_1^6
S_2^6}\eqn\leadingw$$
This superpotential has an approximate U(1) symmetry, broken
by the final two terms:
$$S_2 \rightarrow e^{i b} S_2 ~~~~~~\bar S_2 \rightarrow e^{-ib}
\bar S_2 \eqn\spq$$
This symmetry can play the role of a Peccei-Quinn symmetry.
If $S_2$ and $\bar S_2$ have soft-breaking mass terms of the
correct sign, these fields will acquire expectation
values of order $M_{INT}$, breaking the symmetry spontaneously
(in this model, $S_1$ and $\bar S_1$ obtain larger expectation
values).

The authors of ref. \ross\ estimated the $\theta$ which would
arise in this model by considering the explicit breaking terms
in the superpotential, above, as well as soft-breaking terms
of the type $A m_{3/2} W$.  It is easy to see that these
lead to quite a small $\theta$.   However, a generic supergravity
model also leads to soft supersymmetry-breaking
terms for scalar fields,
$\phi$, of the type $m_{3/2}^2\phi^{*~n}\phi^m$.
In the present case, this allows the operator:
$${m_{3/2}^2 \over M_p^2} S_1^* \bar S_1 S_2 \bar S_2^*
\eqn\badoperator$$
This breaks the Peccei-Quinn symmetry.  It
is a huge term on the scale of axion physics; it gives,
for example, a contribution to the axion mass of order $MeV$'s!

Clearly we can improve the situation if we consider a different
symmetry.  For example,
$$S_1 \rightarrow -\alpha S_1~~~~~\bar S_1 \rightarrow \alpha^2 \bar S_1
{}~~~~~S_2 \rightarrow - S_2~~~~~\bar S_2 \rightarrow \bar S_2
\eqn\newsymmetry$$
again gives a lagrangian which admits a Peccei-Quinn symmetry.
Now, however, the leading symmetry-violating operators are things
like $\bar S_1 S_1^{*~2} S_2^2 \bar S_2^{*}$. This leads to a
$\theta$ which is perhaps barely small enough.

\REF\dineleigh{M. Dine and R.G. Leigh, SCIPP preprint in
preparation.}
We  will not attempt, here, to consider all aspects of the
phenomenology of these models; suffice it to say that it
does appear to be possible to build realistic models along
these lines.  One can debate how
reasonable -- or contrived -- this solution appears to be.
As we will explain elsewhere, it is probably not much better
or worse than is required to obtain a Nelson-Barr type
solution of the problem.\refmark{\dineleigh}  The main
difference is that in the Nelson-Barr case, it is not
necessary to suppress operators of such high dimension.
However, as illustrated by the examples
above, rather simple discrete symmetries can
accomplish this.

\chapter{Some Wild Speculations on Strong CP and
the Cosmological Constant}

I would like to conclude by describing some wilder ideas about
the strong CP problem and the cosmological constant.
\REF\bds{T. Banks, M. Dine,
and N. Seiberg, Phys. Lett. {\bf 273B} (1991)
105.}
These are associated with what T. Banks, N. Seiberg and I have
dubbed ``irrational axions."\refmark{\bds}  Such axions do not arise in
conventional theories.  It is tempting to think that they
might arise in string theory, but we have not found an example
of this phenomenon.  The basic ideas are very simple.  Consider
first the strong CP problem.
Suppose that in
addition to QCD, one has an additional strongly interacting gauge
group; we will refer to this as axicolor, $QAD$.  In addition,
suppose one has a single Peccei-Quinn symmetry, and
$a$ is the associated axion.  The couplings of the axion to the two
gauge groups are written
$${1 \over f_1}{g_3^2 \over 16 \pi^2} \int d^4x a F \tilde F
+ {1 \over f_2}{\tilde g^2 \over 16 \pi^2} \int d^4 x a G \tilde G
\eqn\irrational$$
where $G$ refers to the axicolor gauge fields, and we assume
$\Lambda_{QAD}>>\Lambda_{QCD}$.  Then the axion acquires
its mass principally from $ACD$ dynamics; one expects
$$m_a^2 \sim {\Lambda_{QAD}^4 \over f_2^2}\eqn\bigaxionmass$$
Ordinarily, such an axion would have nothing to do with
the solution of the strong $CP$ problem.  But suppose $f_1 \over f_2$ is
irrational.  The axion potential in this theory is something of the
form
$$V= \Lambda_{ACD}^4 cos(a/f_1) + \Lambda_{QCD}^4 cos(a/f_2 + \theta)
\eqn\axionpotential$$
Now since $f_1 /f_2$ is irrational, one can always
find integers $n_1$ and $n_2$ such that
$$a\approx 2\pi n_2 f_2 \approx (2 \pi n_1 - \theta_2) f_1
\eqn\aminima$$
with  {\it arbitrary} accuracy.  Thus in this theory
there exist ground states with arbitarily small $\theta_{QCD}$.
These states are stable on cosmological scales, but they are
also rare; about $1$ in $10^9$ local minima of the potential
has such small $\theta$.  Thus, in this model the $\theta$
problem is solved without a light axion; the price one pays
is cosmological:  why do we find ourselves in a suitable state?

This problem can (almost) be solved if we invoke the
anthropic principle, in the ``weak sense" discussed
by Weinberg.\refmark{\weinberg}
Suppose that the cosmological constant of the theory is adjusted
so that it vanishes as $\theta \rightarrow 0$.  If this
is the case, formation of galaxies requires that the cosmological
constant not be larger than about $1000$ times the present
experimental limit.  This gives $\theta$ far smaller than
$10^{-9}$.  We still have a factor of $1000$ in cosmological
constant to explain.  Actually, it is not quite as bad as that;
since the energy goes as $\theta^2$, we are really only out
by a factor of $30$.

We have already seen that there are many ways to solve the
strong CP problem, so one more, which we don't (yet) know
how to get from an underlying microscopic theory might not
seem that exciting.  However, there is another fine tuning
problem which we might like to solve without a light particle:
the cosmological constant problem.  This may also be possible
\REF\abbott{L. Abbott, Phys. Lett. {\bf 150B} (1985) 427.}
with irrational axions.  The idea we will describe here
bears some resemblance to a suggestion of Abbott.\refmark{\abbott}

Suppose, in addition to QCD, the underlying theory has two
strong gauge groups, with scales $\Lambda_1 \approx \Lambda_2$.
Suppose the ``irrational axion" couples to these, in the
same irrational way as before.  Suppose also that the bare
cosmological constant, $\Lambda_o^4$, satisfies $\Lambda_o^4
< \Lambda_1^4$.  With supersymmetry, one can show that these
conditons can be natural.  Now the
axion potential is a sum of three terms:
$$V(a) = \Lambda_o^4 + \Lambda_1^4 cos(a /f_1)
+ \Lambda_2^4 cos (a/f_2 + \theta_2) \eqn\cosmopotential$$
In this potential
there exist vacua with arbitrarily small values of the
cosmological constant.  They are even more rare than in our
previous example; for example, if $\Lambda_1 \sim 10^{10}$,
then only one in about $10^{88}$ vacua are acceptable.
Actually, the situation is even worse because in this case
a typical local minimum with small cosmological constant
is not even approximately stable.  Only a small fraction
are:  for example, if several adjacent minima all have higher
energy, the tunneling rate will be suppressed.  Again, we
can suppose that in an inflationary universe, some worlds
like our own were created, and try and invoke the anthropic
principle.  Again, we are off by a factor of $1000$.

However, I don't believe this problem is so severe; for
example, somewhat stronger verions of the anthropic
principle might save the day.  If
we had examples of such irrational axions,
they might well solve the cosmological constant problem.

\bigskip
{\bf Acknowledgements}
I wish to thank my collaborators T. Banks, R. Leigh, D. MacIntire
and N. Seiberg for their insights, and Ann Nelson for several
helpful discussions.

\refout
\end

It was confronted quantitatively
by Casas and Ross,\refmark{\ross} who considered the possibility that
axions might arise by accident as a consequence of discrete
symmetries in string theory.  (The possibility of such accidental
axions was also considered by Lazarides et al.\refmark{\shafi})
In such theories, the Peccei-Quinn symmetry is violated by
high dimension operators; in a particular example, these authors
argued that $\theta$ would be small enough.
We will assume that the model possesses
a $Z_7$ symmetry under which
$$\Phi_i \rightarrow \alpha^2 \Phi_i~~~~~~\chi_a \rightarrow \alpha^3
\chi_a\eqn\supertrans$$
This rule allow Yukawa couplings of Higgs to ordinary matter fields,
$\chi \Phi \Phi$, while forbidding baryon and lepton-number violating
couplings.{\bf check these statements****}
{}From our perspective, the most important feature of this
model is that all of the lowest-dimension couplings, at tree
level, preserve the global symmetry
$$\Phi_i \rightarrow e^{i \omega} \Phi_i~~~~~~~~\chi \rightarrow
e^{-2 i \omega} \chi \eqn\pqtranstwo$$


We will assume that supersymmetry is broken in this theory as in
\REF\nilles{P. Nilles, Phys. Rep. {\bf 110} (1984) 1.}
standard ``hidden sector" supergravity models.\refmark{\nilles}
This gives rise, in the ``visible sector" of the theory,
to explicit supersymmetry breaking couplings.
They are generically of two types:  $m_{3/2} \phi \dots \phi$
and $m_{3/2}^2 \phi^* \dots \phi^* \phi \dots \phi$.
Some of these couplings will lead to expectation values
for scalars.  Generically, these will be of order the scale
\REF\nappietal{M. Dine, V. Kaplunovsky, M. Mangano, C. Nappi and
N. Seiberg, Nucl. Phys. {\bf B259} (1985) 549.}
$f_a$ described above, or larger.  Now we must ask what are the
lowest dimension operators consistent with gauge invariance
which break the symmetry.  In the superpotential, the lowest
dimension terms one can write are of the form
$\Phi^6 \chi^3$ (dimension 10).  Among soft-breaking
terms, the worst are {\bf the authors Ross and Shafi et al
must be dealt with much more nicely in the text}
{\bf if doublets in 27, not $\bar 27$, worst operators will
be as before, since we considered phi* phi type operators
anyway}
of the type
$$m_{3/2}^2 \Phi^{*2} \Phi \eqn\worstoperator$$
Because of the factors of $m_{3/2}^2$, this operator is effectively
of dimension $9$, and is barely ok if
is of order $f_a \sim 10^{10}$.  More elaborate symmetries,
clearly can yield still further suppression.

This model suffers from a number of deficiencies, some of
which are easily remedied.  First, as pointed out by
Lazarides et al\refmark{\shafi}, at low energies,
models of this type possess additional, approximate symmetries
involving separate rotations of Higgs fields, and these give
unwanted light particles.  This problem can be avoided,
as they note, by considering theories in which the theory
below the Planck scale has a gauge group of rank 5.
(One construction of rank 5 models is described in
\REF\miracles{M. Dine and N. Seiberg, Nucl. Phys. {\bf B306}
(1988) 137.}
ref. \miracles.)
Needless to say, the final model
seems rather elaborate (contrived?).  The smallness of
$\theta$ here seems to be an accident.  But the model also
has some virtues:  the same symmetry forbids proton decay,
and the axion decay constant is naturally of the right
order from a cosmological/astrophysical viewpoint.  It would
be of interest to examine real string models with symmetries of
this type.  One might hope, for example, that the same discrete symmetries
which give rise to the Peccei-Quinn symmetry would significantly
constrain the fermion mass matrix.

One can compare this with what might be required to obtain
a solution of the strong CP problem along the lines of
Nelson and Barr.  It turns out to be possible to realize this
solution as well in the context of models with intermediate scale
breaking.\refmark{\dineleigh}
In this framework, again, it is necessary to postulate
discrete symmetries, but
it is only necessary to suppress
a set of operators of rather low dimension.  Perhaps, then
the Barr-Nelson solution is the more plausible one in this
framework.
\refout
\end